\documentclass[12pt, preprint]{aastex}
\begin{document}
\title{Origin of  Ultrahigh Energy Galactic Cosmic Rays:  The Isotropy Problem }
\author{
 Martin Pohl\altaffilmark{1,2},
David Eichler\altaffilmark{3}}
\altaffiltext{1}{Institut f\"ur Physik und Astronomie, Universit\"at
Potsdam, 14476 Potsdam-Golm, Germany}
\altaffiltext{2}{DESY, 15738 Zeuthen, Germany}
\altaffiltext{3}{Physics Department, Ben-Gurion University, Be'er-Sheva 84105, Israel}

\begin{abstract}
We study the propagation of ultra-high-energy cosmic rays (UHECR) in the Galaxy, concentrating
on the energy range below the ankle in the spectrum at 4 EeV. A Monte-Carlo method, based on analytical solutions to the
time-dependent diffusion problem, is used to account for intermittency by placing sources at random locations.
Assuming a source population that scales with baryon mass density or star formation (e.g. long GRB), we
derive constraints arising from intermittency and the observational limits
on the composition and anisotropy. It is shown that the composition and anisotropy at $10^{18}$~eV
are difficult to reproduce and require that either a) the particle mean free path is much smaller than a gyroradius,  implying the escape time is very long,
b) the composition is heavier than suggested by recent Auger data, c) the ultrahigh-energy sub-ankle component is mostly extragalactic, or d) we are living in a
rare lull in the ultrahigh-energy cosmic-ray production, and the current ultrahigh-energy
cosmic-ray intensity is far below the Galactic time average. We therefore recommend a strong
observational focus on determining the UHECR composition around $10^{18}$~eV.  
\end{abstract}

\keywords{Galaxy, cosmic rays, gamma-ray bursts}

\section{Introduction}
An open problem in cosmic-ray (CR) astrophysics is at what energy we observe
the transition from a Galactic to an extragalactic CR origin, and what sorts of sources,
in and out of our own Galaxy, are responsible for each population.
It is widely suspected that the abrupt flattening of the spectrum at $E=4\times 10^{18}$ eV,
the so-called "ankle" of the CR spectrum, indicates the Galactic to extragalactic
transition, but the abrupt increase in loss rate above the ankle due to pair production
with the microwave background may also play a role.  It is generally recognized
that blast waves from supernova remnants cannot accelerate CR protons beyond energies of
about $10^{14.5}$ eV \citep{1983A&A...125..249L}, so they could not reach the CR ankle,
unless the Galactic
magnetic field is amplified by the blast wave by many orders of magnitude.
This motivated \citet{le93} to propose that UHECR are produced by
GRB that occur in our Galaxy.

This is the third paper of a series, earlier publications of which demonstrated that a) long
GRB likely have insufficient power to provide UHECR at energies above the observed
ankle at $4\times 10^{18}$~eV \citep{egp10}, unless their CR power is much higher than
their $\gamma$-ray power, and b)
at energies below the ankle the contribution from Galactic GRB should on average be stronger
by about 3 orders of magnitude than that of extragalactic GRB (as the latter would imply that
extragalactic space is filled uniformly), thus further constraining models
involving a GRB origin of trans-ankle particles \citep[henceforth referred to as Paper II]{ep11}.

In this paper we study the time-dependent diffusive transport of UHECR in the Galaxy using the
method of Monte-Carlo to account for the unknown location and explosion time of GRB or other
sources with similar population statistics. For simplicity, we shall call the sources GRBs
throughout the paper, but it is to be
understood that the study applies to other classes of UHECR sources
as well (e.g. $\gamma$-ray silent hypernovae). Our approach permits us to quantitatively account for intermittency
effects in the local UHECR spectrum and thus goes beyond the scope of earlier publications
\citep{le93,2004APh....21..125W,2010PhRvL.105i1101C}. Faced with the observed energy spectrum
and low limits on anisotropy recently set by Auger and other experiments, we calculate spectra
and anisotropy that would be expected as a function of model  parameters. The ceiling on
anisotropy, of order 1 percent or less below $10^{18}$ eV, is interesting given the rather
large mean free paths of such energetic cosmic rays.

We  assume the propagation in the Galaxy of cosmic rays at energies $10^{15}$~eV to
$10^{18}$~eV can be accurately described as isotropic diffusion.
The framework of isotropic diffusion
requires that the particle Larmor radii be considerably smaller than the dimensions over which the diffusion is considered.  If this were not the case, our conclusion that isotropy is hard to understand would be strengthened, as there would be little chance of isotropizing UHECR emerging from an anisotropic source distribution.
Likewise, the particle mean free path, $\lambda_{\rm mfp}$, should be much smaller
than a few kpc, the typical distance between
the solar system and a GRB in the Galaxy. The Larmor radius of a $Z=1$ particle in a $10\ {\rm
\mu G}$ field is used as a scale in this paper. It reaches $\sim 100$~pc at $10^{18}$~eV, and 
therefore the first approximation should
hold for UHECRs of any composition below $\approx 10^{18}$~eV. The second approximation
requires that $\lambda_{\rm mfp}$ be within a factor of $\sim 10$ of the Larmor radius.
\citet{2009ApJ...693.1275A} have studied the transition from the rectilinear regime to the 
diffusive regime of particle propagation, concluding that diffusion models are a good approximation
if the source distance is $L>6\,\lambda_{\rm mfp}$. In the models presented in Section \ref{sec3}, for protons at $E=4$~EeV $\lambda_{\rm mfp}\leq 500$~pc, and smaller by a factor $1/Z$
for heavy nuclei, whereas the GRB rate is so low that typically $L\gtrsim 5$~kpc. As we shall demonstrate,
reproducing the observed low anisotropy in the EeV band requires a mean free path small enough to
maintain validity of the diffusion approach, at least for GRB.

Isotropic diffusion requires that the magnetic field is weakly ordered. Measurements
of Faraday rotation indicate that the strength of the ordered field is about $2\,{\rm \mu G}$,
whereas the random field has an amplitude of $\sim5\,{\rm \mu G}$ on scales of 10 -- 100 pc
\citep[and references therein]{2009IAUS..259..455H}. Note that UHECR sample the turbulent field in the spectral band where they are strongest, 
on the scales of energy injection by supernovae and stellar winds.

To evaluate the level of systematic uncertainties in our model description, we explore various
geometric forms of the propagation volume of UHECR in the Galaxy. We find that a disk-like geometry,
which appears more likely to be accurate than the assumption of spherical symmetry, renders
the observational constraints on anisotropy and composition more difficult to meet, and resolution of this problem could lead to interesting or unconventional conclusions.

\section{Propagation of UHECRs from Galactic GRB}
\subsection{Spherical symmetry}
Treating energy losses as a catastrophic loss term with loss time $T$,
assuming the particles are instantaneously released at time $t_0=0$
by a single point source, and assuming a large cosmic-ray halo, the problems is
spherically symmetric with respect to the GRB, and
the differential density of UHECR obeys a continuity equation that can be written as
\begin{equation}
\frac{\partial N}{\partial t} + \frac{N}{T} - \frac{1}{r^2}\frac{\partial}{\partial r}
\left(r^2\,D\,\frac{\partial N}{\partial r}\right)=Q(E)\,\delta (t)\,
\frac{\delta (r)}{4\pi\,r^2}
\label{diff-eq}
\end{equation}
Here, $D=c\,\lambda_{\rm mfp}/3$ is the spatial diffusion coefficient, $r$ is the distance from
the GRB, and $Q(E)$ is the differential production rate of UHECR in a GRB. 
We assume that we are located outside of the dissipation region of the GRB, implying that the plasma no longer streams with significant velocity. If we were, the Compton-Getting anisotropy would be
prohibitively large. Once the GRB outflow turns non-relativistic, the further dynamical evolution 
is not unlike that of a supernova remnant, and we can expect the GRB remnant to be of order 100 pc in
size, small enough to render ii highly unlikely that we reside inside. The high-energy particles escaping from a GRB can initially have a rather directional distribution which leads to well-known instabilities \citep[e.g.][]{1999ApJ...526..697M,2000A&A...354..395P} that provide for a rapid isotropization of the particles. We assume that this isotropization is achieved on a scale small compared with the average distance to GRBs, and hence we can approximate it as being instantaneous.
 
Note that
equation~\ref{diff-eq} contains the particle energy, $E$, only parametrically, it is not a
variable. The catastrophic loss term permits using Dirichlet boundary conditions
$N(r=\infty)=0$.

Using standard methods \citep{1962SvA.....6..317K,1964ocr..book.....G} one derives
a general solution to the transport equation,
\begin{equation}
N(r,t,E)= \exp\left(\frac{-t}{T}\right)\,\frac{\Theta(t)}
{\left(4\pi\,D\,t\right)^{3/2}}\,Q(E)\,\exp\left(-\frac{r^2}{4\,D\,t}\right)
\label{n-eq}
\end{equation}
If more than one GRB contribute to the local UHECR flux at any time, their individual
contributions must be calculated using equation~\ref{n-eq} and then summed.

The anisotropy in the UHECR intensity mainly arises from the diffusive
flux \citep[e.g.][]{1989ApJ...336..243S} and is of the order
\begin{equation}
\delta= \frac{I_{\rm max}-I_{\rm min}}{I_{\rm max}+I_{\rm min}}\simeq
\lambda_{\rm mfp}\,\frac{1}{N_{\rm tot}}\,
\left|\vec\nabla N_{\rm tot}\right|
\label{aniso}
\end{equation}
where $N_{\rm tot}$ is the total UHECR spectrum after summing the contributions of
all relevant GRB. If there was only one GRB, then the anisotropy would be
\begin{equation}
\delta\simeq \frac{3\,r}{2\,c\,t}
\label{aniso1}
\end{equation}
and therefore, setting for example $r=r_\odot$, a given anisotropy at the solar circle, $\delta$,
requires that the age of the UHECR population, $t$, obey
\begin{equation}
t\simeq \frac{3\,r_\odot}{2\,c\,\delta}= (4\cdot 10^6\ {\rm yr})\,
\left(\frac{\delta}{10^{-2}}\right)^{-1}
\label{aniso2}
\end{equation}
At that time, the Galactic UHECR fill a sphere of radius
\begin{equation}
R\simeq \sqrt{4\,D\,t}\simeq (10\ {\rm kpc})\,
\left(\frac{\lambda_{\rm mfp}}{100\ {\rm pc}}\right)^{1/2}
\,\left(\frac{\delta}{10^{-2}}\right)^{-1/2}
\label{aniso3}
\end{equation}
To be noted from the approximation represented by equations \ref{aniso1} and
\ref{aniso3} is that the UHECR flux at the
solar circle depends on the mean free path, but the anisotropy does not
\citep[e.g.][]{2004APh....21..125W}.

The above conclusion is highly significant. It
means that at increasingly high energies the anisotropy may be reduced, given a
sufficiently long time since the last point-source injection, to
within the tight observational constraints even at energies high enough
that the mean free path is a significant fraction of the geometric
scale of the Galaxy, as observations indicate. We therefore investigate
how the solution generalizes to arbitrary radial dependence of
the diffusion coefficient.

Consider, then, equation \ref{aniso1} with a radially dependent diffusion coefficient
$D=D_or^j$, with $j>0$.
 \begin{equation}
\frac{\partial N}{\partial t} + \frac{N}{T} - \frac{1}{r^2}\frac{\partial}{\partial r}
\left(r^2\,D_o r^j\,\frac{\partial N}{\partial r}\right)=Q(E)\,\delta (t)\,
\frac{\delta (r-r_o)}{4\pi\,r^2}
\label{rdep}
\end{equation}
Note that we choose a finite radius of injection, $r_o$, in a way
that preserves spherical symmetry, so that the diffusion coefficient
at the injection site is finite.
For $j<2$,  a finite boundary condition at  the origin and a vanishing
boundary condition at $r=\infty$, the solution at large times
(or $r\gg r_0$) approaches
 \begin{equation}
N(r,t,E)\simeq \exp\left(\frac{-t}{T}\right)\,\frac{\Theta(t)}
{\left(4\pi\,D_o r_o^j t\right)^{3/(2-j)}}\,Q(E)\,
\exp\left(-\frac{r^2}{(j-2)^2\,D_o r^j\,t}\right).
\label{n-eq2}
\end{equation}
By comparison with Equation \ref{n-eq}, the case of constant
diffusion coefficient, the solution falls off more quickly after
reaching a peak flux at $r^2=3\,(2-j)\,D_0\,r^j\,t$,
whereas the anisotropy, when estimated as in Equation \ref{aniso}, is
\begin{equation}
\delta = \frac{3\,r}{(2-j)\,c\,t} ,
\end{equation}
(which differs from the expression in Equation \ref{aniso1}  by the
numerical factor of order unity ($1-\frac{j}{2}$)). That the anisotropy has similar time
dependence, while the decline in intensity has steeper time dependence has further
significance:  It means
that to reduce the anisotropy with time at the observer's location, a {\it larger}
reduction in intensity is required for positive j than for $j=0$ (the case of spatially
homogeneous diffusion). If we
were to attempt accommodating the low observed UHECR anisotropy with an anisotropic
source distribution by invoking an intermittent source of UHECR followed by a period
of zero source activity, we would have to pay a larger cost in intensity reduction.
The intensity at Earth at the time of maximum intensity, we would conclude, would be
higher than the current intensity by an even larger factor than in the case
of homogeneous diffusion.

When $j>2$  the exponential factor does not vanish at $r=\infty$,
and the correct Green's function at large time is

 \begin{equation}
N(r,t,E)= \exp\left(\frac{-t}{T}\right)\,\frac{\Theta(t)}
{\left(4\pi\,D_o r_o^j t\right)^{(1-2j)/(2-j)}}\,Q(E)\,(r/r_o)^{-j-1}
\exp\left(-\frac{r^2}{(j-2)^2\,D_o r^j\,t}\right).
\label{n-eq3}
\end{equation}

In this case, the scale height of N at radius r, $(d\ln N/dr)^{-1}$, is of order r,
even at large t, and the anisotropy does not become arbitrarily small in the limit
of large t. If the magnetic field of the Galaxy is dragged out by a Galactic wind,
it is quite reasonable to suppose that the magnetic field decreases as $1/r^2$, and
that the gyroradius and mean free path increase at least as fast as $r^2$. It is
therefore not clear whether the small observational upper limit on anisotropy
can be accommodated with a
realistic magnetic field geometry, no matter how intermittent the sources. Similar
remarks would apply if convection were
significant as compared to diffusion.  Moreover, we have not yet taken into account
the fact that a single intermittent source of cosmic rays would be localized in angle
as well as radius, which would further increase the anisotropy.

Below we reconsider
the anisotropy question assuming a free-escape boundary up to several kpc above the Galactic
disk following the consideration of several other issues. The free-escape boundary
models the effect of a very strong dependence of $D$ with $r$ (e.g. due to a weaker
magnetic field) and/or the outward convective
effects of a Galactic wind or of Galactic chimneys beyond some region. We show that the
same principle applies as demonstrated above in the spherically-symmetric case with
spatially varying coefficient:  that the leakage out the faces of the disk greatly
reduces the intensity of the UHECR as they isotropize relative to the observer;
in fact, while the anisotropy declines algebraically with time, as in the previous
examples, the intensity declines exponentially, so the isotropization carries a large
cost in intensity reduction. This poses constraints on any theory of their origin from
sources that are distributed anisotropically in the observer's sky.

\subsubsection{Heavier nuclei}\label{heavy}
Heavier nuclei have a
smaller rigidity than protons at the same total energy, $R\propto E/Z$. The mean free path of an
ultra-high-energy particle should only depend on the rigidity, and in the absence
of energy losses a nucleus of charge $Z$ and energy $E_Z$ should behave like a
proton of energy $E=E_Z/Z$. Thus equation \ref{n-eq} also describes the distribution
of heavy nuclei in the Galaxy, provided the appropriate scaling is applied to
the energy and the source rate.

\subsection{Diffusion in disk geometry}
Cosmic-ray studies at lower energies indicate that a few kpc out into the halo
diffusive transport become much faster, and particles efficiently escape from the Galaxy.
For a cosmic-ray source near, e.g., the Galactic Center we therefore expect deviations
from spherical symmetry, and a thin-disk geometry appears better suited to describe
the transport of cosmic rays.
The cosmic-ray continuity equation (\ref{diff-eq}) then changes to
\begin{equation}
\frac{\partial N}{\partial t} + \frac{N}{T} -
\frac{\partial}{\partial x}\left(D\,\frac{\partial N}{\partial x}\right)
-\frac{\partial}{\partial y}\left(D\,\frac{\partial N}{\partial y}\right)
-\frac{\partial}{\partial z}\left(D\,\frac{\partial N}{\partial z}\right)
=Q(E)\,\delta (z)\,\delta (t)\,q(x,y)
\label{diff-eq-cyl}
\end{equation}
where the halo boundary can be incorporated through boundary conditions $N(z=\pm H)=0$.
Integrating over $z$ yields
\begin{equation}
\frac{\partial M}{\partial t} + \frac{M}{T} -
\frac{\partial}{\partial x}\left(D\,\frac{\partial M}{\partial x}\right)
-\frac{\partial}{\partial y}\left(D\,\frac{\partial M}{\partial y}\right)
-\left(D\,\frac{\partial N}{\partial z}\right)\bigg\vert_{-H}^H
=Q(E)\,\delta (t)\,q(x,y)
\label{diff-eq-z}
\end{equation}
where
\begin{equation}
M=\int_{-H}^H dz\ N
\end{equation}

\begin{figure}[t]
\includegraphics[width=0.7\columnwidth, keepaspectratio]{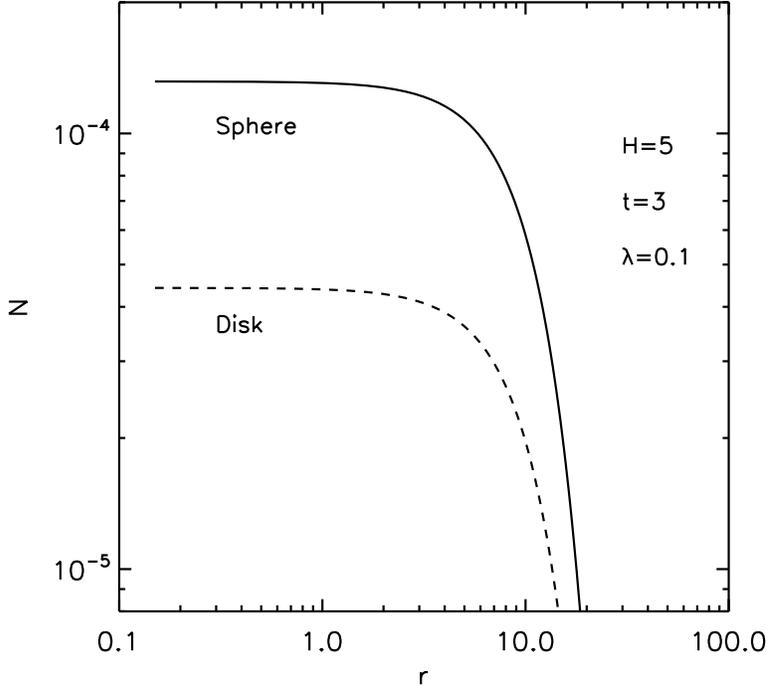}
\caption{CR density as function of r ($\rho)$ for isotropic
diffusion (solid line) and disk geometry (dashed). The parameters are $t=3$~Myr,
$H=5$~kpc, and $\lambda=0.1$~kpc. \label{spec}}
\end{figure}
Instead of a computationally expensive full solution \citep{2005ApJ...619..314B},
we use a steady-state solution to estimate the relation between $M$ and the mid-plane
cosmic-ray density, $N_0$, as well as turn the diffusive flux at the halo boundaries into a
simple catastrophic loss term. In the energy band of interest, escape is the dominant loss
process of cosmic rays in the Galaxy, and in the steady state we expect their density to follow
\begin{equation}
N(z)=N_0\,\left(1-\frac{\vert z\vert}{H}\right) ,
\end{equation}
implying
\begin{equation}
M=H\,N_0\quad {\rm and}\ \
-\left(D\,\frac{\partial N}{\partial z}\right)\bigg\vert_{-H}^H
=\frac{2\,D}{H^2}\,M .
\end{equation}
The diffusive escape can thus be described using a loss time
\begin{equation}
\tau_{\rm esc}=\frac{H^2}{2\,D}\simeq (1.2\cdot 10^6\ {\rm yr})\,
\left(\frac{H}{5\ \rm kpc}\right)^2\,
\left(\frac{\lambda_{\rm mfp}}{0.1\ \rm kpc}\right)^{-1}
\end{equation}
which is far shorter than the energy-loss time, $T$, which is 
$\gtrsim 10^9$ years for photo-secondary production in the Galaxy \citep{1988SvAL...14....1B} and
$\sim 10^8$ years for secondary production in p-ISM collisions \citep[with average density
$n_{\rm ISM}<0.2$~atoms /cm$^3$]{2007APh....27..429H}, 
and therefore energy losses can be ignored henceforth.
Ignoring variations in the diffusion coefficient within the Galactic plane,
the problem only depends on the in-plane distance between source (GRB) and observer, $\rho$,
and can be recast as 2-D diffusion equation for the mid-plane cosmic-ray density
around a point source.
\begin{equation}
\frac{\partial N_0}{\partial t} + \frac{N_0}{\tau_{\rm esc}} - \frac{1}{\rho}\,
\frac{\partial}{\partial \rho}\left(\rho\,D\,\frac{\partial N_0}{\partial \rho}\right)
=Q(E)\,\delta (t)\,\frac{\delta (\rho)}{2\pi\,\rho\,H}
\label{diff-eq-neu}
\end{equation}
with solution
\begin{equation}
N_0(\rho,t,E)= \exp\left(-\frac{t}{\tau_{\rm esc}}\right)\,\frac{\Theta(t)}
{4\pi\,D\,t}\,\frac{Q(E)}{H}\,\exp\left(-\frac{\rho^2}{4\,D\,t}\right)
\label{n-eq-neu}
\end{equation}
The anisotropy in the case of a single GRB is the same as in equation \ref{aniso1},
using $\rho$ in lieu of $r$. The ratio between the cosmic-ray density calculated for
a disk geometry and that for spherical symmetry is
\begin{equation}
R=\sqrt{2\pi\,\frac{t}{\tau_{\rm esc}}}\,\exp\left(-\frac{t}{\tau_{\rm esc}}\right)
\end{equation}
which, as expected, becomes small if $t\gg \tau_{\rm esc}$. A comparison of the
radial dependence of the particle density for spherical symmetry as opposed to
disk geometry is shown in Figure \ref{spec}. To be noted is that the radial distribution
is unchanged, the only difference is seen in the normalization.

\subsection{Discreteness Anisotropy}

The above discussion considered anisotropy from a single source in conditions of high
intermittency. When many sources contribute, the anisotropy is reduced by a factor of
order the square root of the number of contributing sources.  We can formally define
intermittency for the case of a disk geometry with absorbing boundaries as follows:
The CR escape time from the disk, $\tau_{\rm esc}$, if the CR are produced in the Galactic plane,
is $\tau_{\rm esc}=H^2/(2D)=3H^2/(2\lambda_{\rm mfp} c)$.
Here $D$, the CR diffusion coefficient, is written as $\lambda_{\rm mfp} c/3$, where
$\lambda_{\rm mfp}$ is the CR mean free path. The intermittency factor $I$ can be defined
to be $1/M_{\rm CR}$ where $M_{\rm CR}$ is the number of CR episodes per diffusion area over one
escape time. Because the CR escape time is the time required to diffuse the height of
the disk, $H$, the diffusion area over one escape time is $\pi H^2$, so
$M_{\rm CR}= q \tau_{\rm esc} H^2/R^2$, where $R$ is the radius of the Galaxy and $q$ is the rate of CR
production events per the entire Galaxy. (For supernovae, $q_{\rm SN}$ is thought to
be about 1/30 yr, and for GRB, $q_{\rm GRB}$ is thought to be somewhere between 30,000
years and 3 million years.) Sources much further away than $H$ are attenuated by escape
and do not contribute significantly. Finally we write $M_{\rm CR}$ as
\begin{equation}
 M_{\rm CR} = \frac{3\,H^4\, q}{2\,R^2\, \lambda_{\rm mfp}\,c}
\end{equation}
and, assuming that the average distance to the sources is of order $H$ and using equation
(3), we write the anisotropy in the case of $M$ contributing sources as
\begin{equation}
 \delta_M \approx \frac{\lambda_{\rm mfp}}{H\,\sqrt{M_{\rm CR}}}=
\left(\frac{\lambda_{\rm mfp}}{H}\right)^{3/2}\,
\frac{R}{H}\, \sqrt\frac{2\,c}{3\,q\,H}.
\end{equation}
The exact anisotropy of course depends on time and numerical factors of order unity. It may be increased by an inhomogeneous spatial distribution of CR sources
in the Galaxy, and therefore in subsequent
sections we shall consider the problem with Monte-Carlo generation of CR events.  It
can be seen however, that $ \delta_M$, at EeV energies, can be considerable. For a
Galactic magnetic field of $10\ {\rm \mu G}$, the gyroradius of a 3-EeV protons is $0.3$~kpc,
and thus the factor $[\lambda_{\rm mfp}/H]^{3/2}$ is more than 1 percent even if
$\lambda_{\rm mfp}$ is but a single gyroradius. The factor $R/H$ is considerably larger
than unity except for a rather thick disk, and $\sqrt\frac{2\,c}{3\,q\,H}$ is at least of
order unity for GRB, and about $10^{-1.5}$ for SN. Unless $H/R$ is chosen to be of order
unity, the diffusion is sub-Bohmian, or the trans-EeV CR are heavy nuclei, the anisotropy
exceeds the experimental limit of 1 percent set by the AUGER experiment
\citep{2011APh....34..627P}. The inverse cubic dependence of $\delta$ on $H$
constrains $H$ to be larger than 3 kpc if SN are the sources of UHE Galactic CR,
and, if GRB are the sources, appears to require them to be intermittent.

The qualitative point is that, if the sources are distributed in the Galactic plane,
then "Olbers paradox" does not apply and close sources contribute as much as distant
ones. This allows the possibility of discrete anisotropy at ultrahigh energies.

\subsection{Intermittency}
\subsubsection{Spectra}
Generally, GRBs in the Galaxy are expected every million years or so, the exact rate depending
on the beaming fraction and the detailed scaling of long GRB with star formation and metallicity
\citep[For a detailed review see][]{2009ARA&A..47..567G}.
Therefore, only a small number of GRB can contribute to the particle
flux at the solar circle, and their relative contribution depends on the
location and explosion time of the GRB. Variations in the local particle
flux must be expected, and neither the particle spectrum from an
individual GRB nor the spectrum calculated for a homogeneous source
distribution are good proxies. To fully account for discreteness of GRBs
in space and time, we can use the method of Monte-Carlo to randomly place
GRBs in the Galaxy with given spatial probability distribution in
galactocentric radius,
\begin{equation}
P(r_{GC})=\frac{2\,r_{GC}}{r_0^2}\, \exp\left(-\frac{r_{GC}^2}{r_0^2}\right)\ ,
\end{equation}
with scale $r_0=5$~kpc, and with given GRB rate, e.g.
\begin{equation}
P(t)=0.5\ {\rm Myr}^{-1} .
\end{equation}
The assumed galactocentric distribution is an approximation to the distributions of baryons
and star formation in the Galaxy, which are reasonable proxies for the distribution of supernovae.
The distribution of long GRB may be skewed toward the outer Galaxy by metallicity-related
selection effects \citep[e.g.][]{2010AJ....140.1557L,2011arXiv1101.4418L}. In this paper
we keep fixed
the galactocentric distribution of UHECR sources. In a forthcoming study we shall
explore the impact of sources in the outer Galaxy, e.g. short GRB.

Each GRB is assumed to inject particles over $10^4$~yr with the same total
energy content and spectrum
\begin{equation}
Q(E)=Q_0\,E^{-s}
\end{equation}
The actual choice of spectral index is irrelevant for the calculation, because
it propagates right into the final expressions for the particle spectra
(cf. equations \ref{n-eq} and \ref{n-eq-neu}). Likewise, it is sufficient to
calculate the variations in the local flux for protons, because nuclei of
charge $Z$ and energy $E_Z$ behave like a proton of energy $E=E_Z/Z$
(cf. section \ref{heavy}).
The halo size is again $H=5$~kpc.

\begin{figure}
\includegraphics[width=0.7\columnwidth, keepaspectratio]{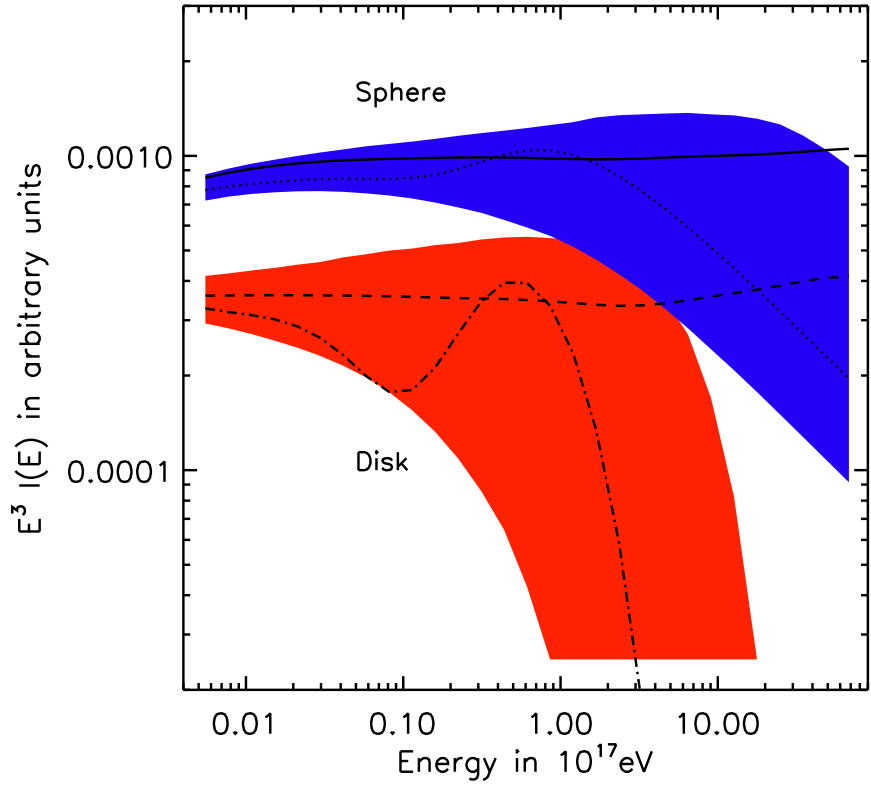}
\caption{Proton spectra at the solar circle expected for a
diffusion coefficient with Bohmian energy scaling (case A). For spherical
geometry, the solid line shows the time-averaged particle spectrum, whereas the
dotted line denotes one randomly selected case. The blue band indicates
the central 68\% containment region for the particle flux at the given energy.
The dashed line, dashed-dotted line, and the red band (partially obscured)
are the same for disk geometry.
\label{specgrb}}
\end{figure}
\begin{figure}
\includegraphics[width=0.7\columnwidth, keepaspectratio]{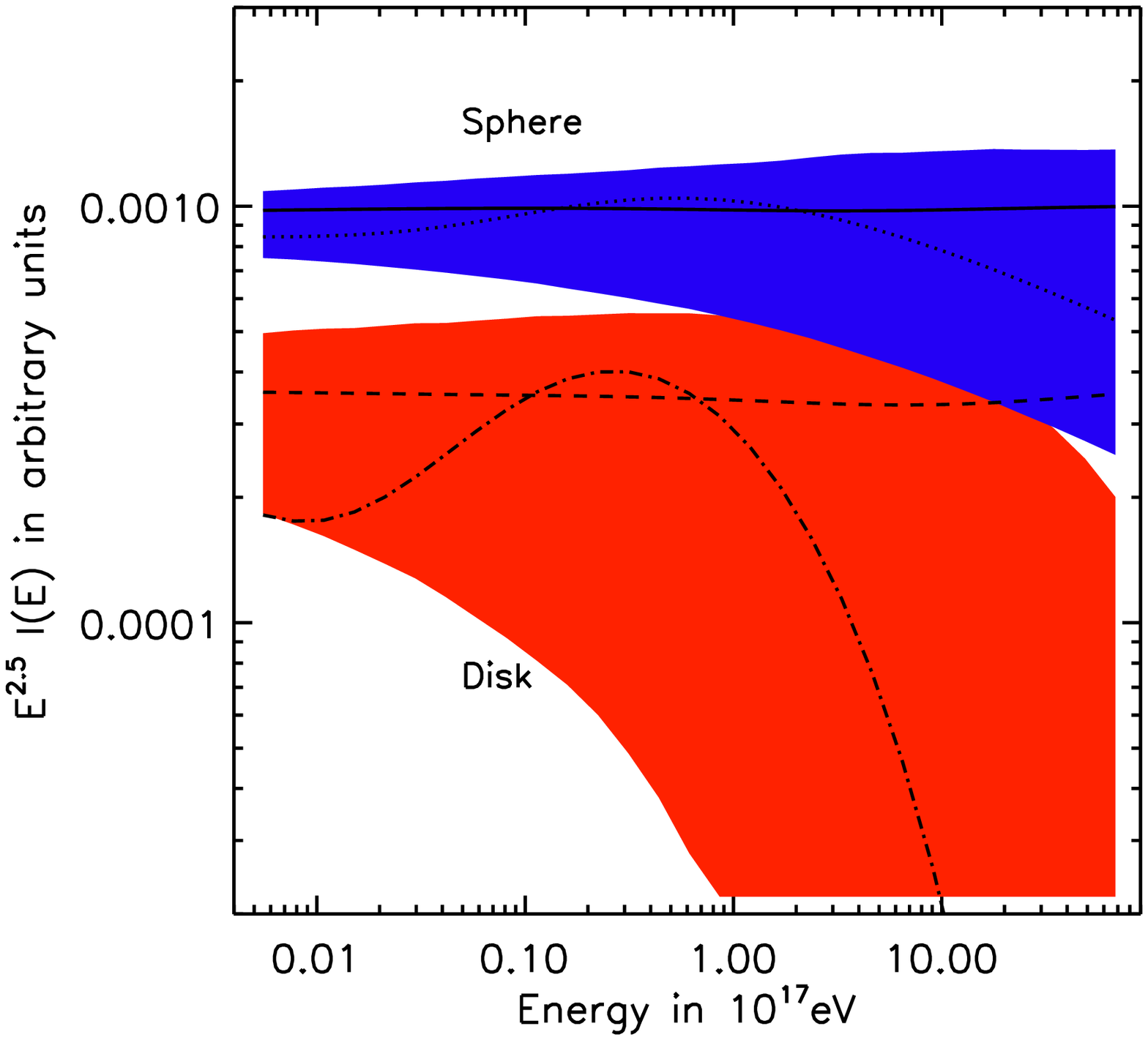}
\caption{Proton spectra at the solar circle expected for a
diffusion coefficient scaling with $\sqrt{E}$ (case B). For spherical
geometry, the solid line shows the time-averaged particle spectrum, whereas the
dotted line is one randomly selected case. The blue band indicates
the central 68\% containment region for the particle flux at the given energy.
The dashed line, dashed-dotted line, and the red band (partially obscured)
are the same for disk geometry. Note that the y-axis is now $E^{2.5}\,I$.
\label{specgrb0.5}}
\end{figure}

Below we show results for two different energy dependences of the diffusion
coefficient, or mean free path.
\begin{eqnarray}
{\rm Case\ A:}\quad& \lambda_{\rm mfp}=10\,r_L\\
{\rm Case\ B:}\quad& \lambda_{\rm mfp}=10\,r_L\,E_{17}^{-0.5}
\label{eq-mfp}\end{eqnarray}
where the Larmor radius is calculated for a proton in a 10-$\mu$G magnetic
field, and $E_{17}=E/(10^{17}\ {\rm eV})$. Case B thus uses a diffusion
coefficient that increases with the square root of the energy, similar to
that conventionally used for Galactic cosmic rays below the knee. Note that for
case A the diffusion approximation breaks down at approximately $10^{18}$~eV
($10^{19}$~eV for case B).

\begin{figure}
\includegraphics[width=0.7\columnwidth, keepaspectratio]{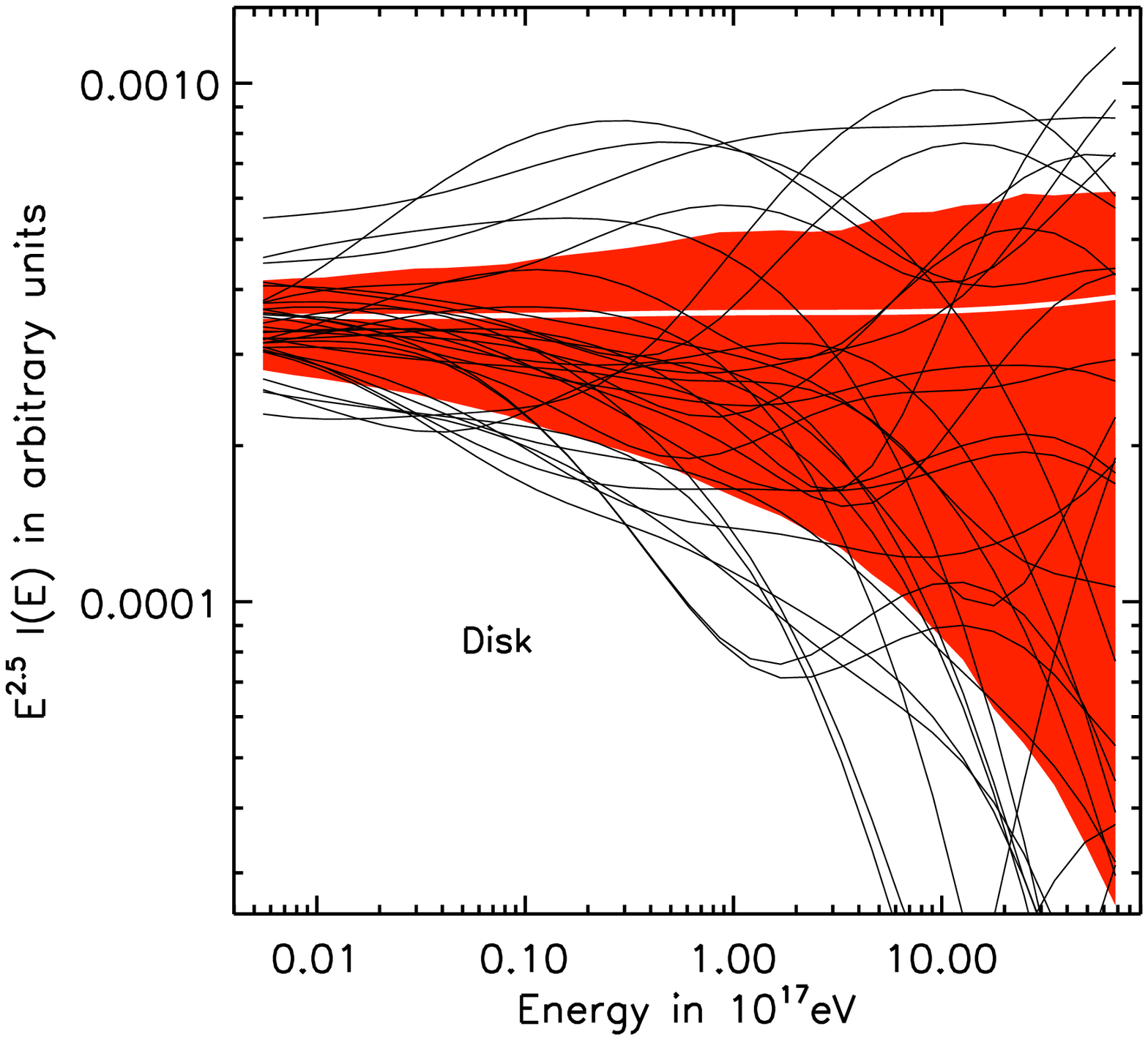}
\caption{Proton spectra at the solar circle expected for a
diffusion coefficient scaling with $\sqrt{E}$ (case B) in disk geometry.
The parameters are as in Figure \ref{specgrb0.5}, except that we have
increased the GRB rate to $5\ {\rm Myr}^{-1}$.
The thin solid lines show 31 randomly selected spectra, whereas the time average is now
given by the thick white line.
\label{specgrb-r5}}
\end{figure}
For both case A and case B we have calculated spectra for $10^4$ random
sets of GRBs. As we demonstrated before, energy losses through
inelastic collisions can be ignored. To reach a quasi-steady state, we need to sample the contributions of all GRBs for a time period much longer than the escape time. We chose
to integrate over 6~Gyr in time, which is long enough at all energies and for all choices of 
$\lambda_{\rm mfp}$ used in this study. Overall, the method is the same as that used to model
the transport of cosmic-ray electrons in the Galaxy
\citep{1998ApJ...507..327P,2003A&A...409..581P}. Results are shown in Figure
\ref{specgrb} for case A, and in Figure \ref{specgrb0.5} for case B, both with injection
index $s=2$. The shaded areas in the Figures indicate the central 68\% containment region
for the particle flux, meaning 16\% of the simulated spectra fall below the
colored range, and another 16\% are higher. 

To be noted from the figures are
\begin{itemize}
\item The energy dependence of the diffusion coefficient influences
the particle spectrum observed at Earth. Features in the observed spectrum could thus
arise from changes in the energy dependence, e.g. from shallow at lower
energies to Bohmian at higher energies, without requiring any structure in the
source spectrum (see also Calvez et al. 2010).
\item The solution for disk geometry is typically a factor 2-3 lower than that
for spherical geometry, reflecting the fact that the free-escape boundary in the halo
introduces particle losses.
\item Intermittency is strong for a GRB rate below 1 per Myr, in particular
for the more realistic disk geometry. In essence, the local UHECR spectrum
from Galactic GRBs is unpredictable if the scattering mean free path exceeds about
100~pc, which for the parameters used here is the case above $10^{17}$~eV for protons, and above
$3\cdot 10^{18}$~eV for iron. Model fits of single-source spectra can thus be very
misleading \citep[cf.][]{2004APh....21..125W}.
\item Particularly interesting is the finding that the time-averaged spectrum
can be outside the 68\% containment region, which arises from the very rare
GRB that are both recent and near. It indicates that it is unlikely that we observe
the time-averaged spectrum. The median of the distribution shows a spectral cut off,
which is therefore the most likely outcome, whereas the time-averaged spectrum does not and thus is
not useful in a comparison with data.
\item The actually expected spectra
display bumps unrelated to both source and propagation physics,
some of which may indeed be observed \citep{2010arXiv1009.4716A}.
The absence of very large bumps in the observed UHECR spectra suggests that
either the mean free path for scattering is smaller than assumed here, or the rate of
cosmic-ray producing GRBs in the Galaxy exceeds 1 per Myr, at which the amplitude of
such bumps becomes smaller (see Figure \ref{specgrb-r5}). Generally,
careful accounting of the statistical fluctuations is mandatory for proper
estimating the local UHECR spectrum from GRBs \citep[cf.][]{2010PhRvL.105i1101C}.
\end{itemize}
\subsubsection{Anisotropy}
If more than one GRB contributes to the local UHECR spectrum, their anisotropy
signal will generally not line up and can thus not be simply averaged.
If all sources reside in the Galactic plane, the entire problem is essentially
two-dimensional. We use an Eddington approximation for the UHECR intensity
from a single GRB,
\begin{equation}
I=I_0 +I_1\,\cos\theta=I_0\,(1+\delta\,\cos\theta) ,
\end{equation}
where $\theta$ is the angle relative to the line of sight from the GRB to us.
The GRB is located at Galactocentric coordinates $(r_{\rm GC},\phi)$, which translates to
distance, $r$, and line-of-sight direction relative to the anticenter, $\psi$, as
\begin{equation}
r^2=64+r_{\rm GC}^2-16\,r_{\rm GC}\,\cos\phi
\end{equation}
and
\begin{equation}
\cos\psi= {{8-r_{\rm GC}\,\cos\phi}\over r}\qquad \sin\psi={{r_{\rm GC}\,\sin\phi}\over r}
\end{equation}
where the distance unit kpc is used.
The local anisotropy can be expressed through projection in anticenter
direction, $\vec e_x$, and in
a direction perpendicular to it, $\vec e_y$.
\begin{equation}
I_x=\delta\,I_0\,\cos\psi\qquad\qquad I_y=\delta\,I_0\,\sin\psi
\end{equation}
After summing over all sources,
\begin{equation}
I_{\rm tot}= \sum I_0\qquad I_{\rm tot,x}= \sum I_x \qquad I_{\rm tot,y}= \sum I_y
\end{equation}
the observed anisotropy is
\begin{equation}
\delta_{\rm tot}={\sqrt{I_{\rm tot,x}^2+I_{\rm tot,y}^2}\over {I_{\rm tot}}}
\end{equation}
and the direction is given by
\begin{equation}
\alpha = \arctan\left({{I_{\rm tot,y}}\over {I_{\rm tot,x}}}\right)
\end{equation}

\begin{figure}
\includegraphics[width=0.7\columnwidth, keepaspectratio]{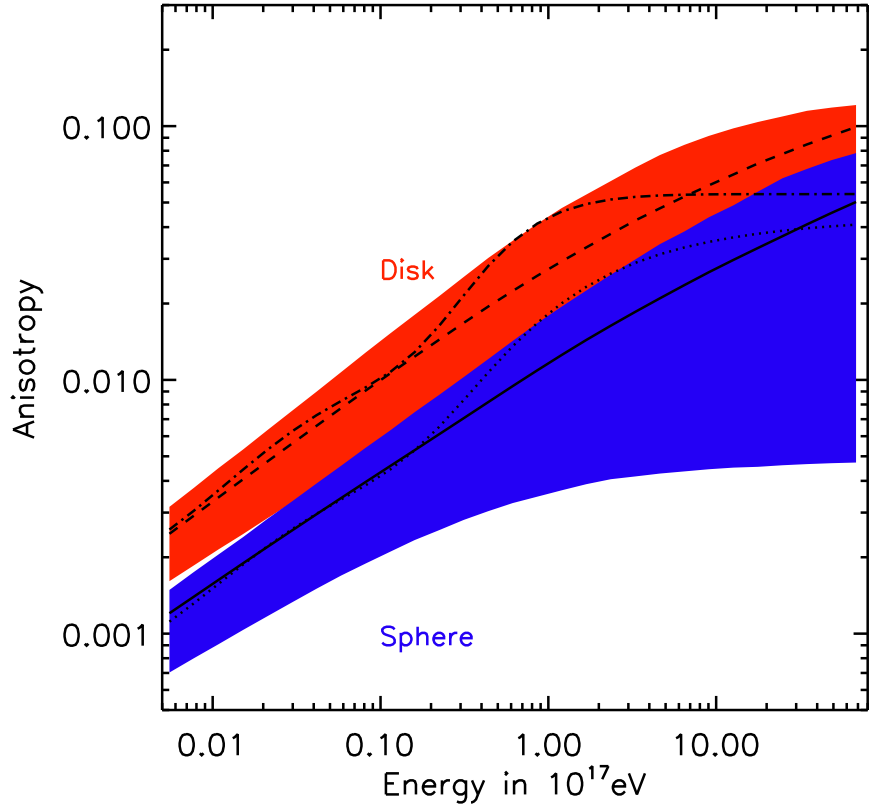}
\caption{The expected anisotropy for the same parameters as in
Figure \ref{specgrb0.5}. For spherical
geometry, the solid line shows the average anisotropy, whereas the
dotted line is one randomly selected case. The blue band indicates
the 68\% containment region for the anisotropy at the given energy.
The dashed line, dashed-dotted line, and the red band (partially obscured)
are the same for a disk geometry.
\label{f-aniso2}}
\end{figure}
\begin{figure}
\includegraphics[width=0.7\columnwidth, keepaspectratio]{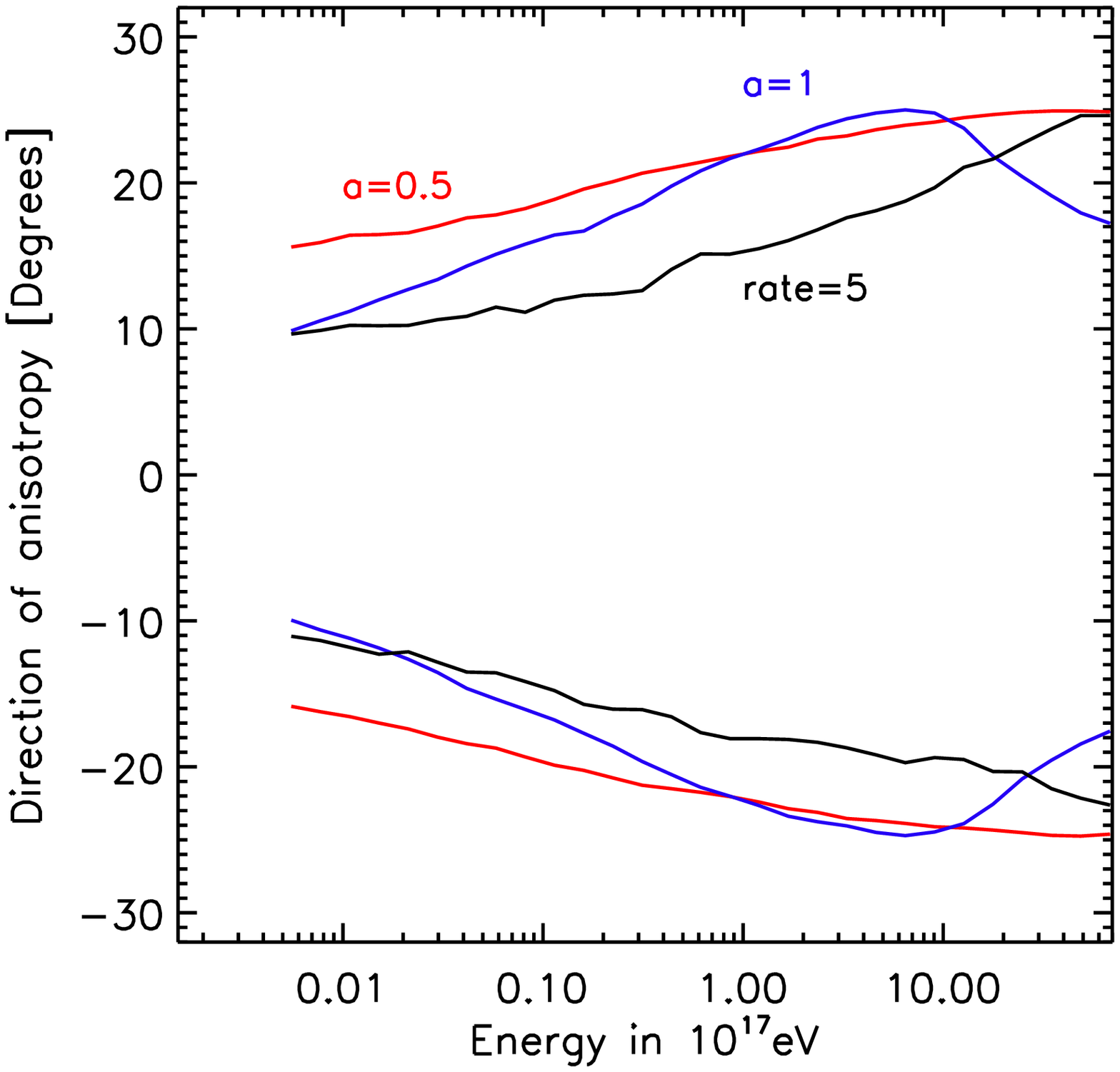}
\caption{The distribution of anisotropy directions
relative to the anticenter for disk geometry. The red and blue lines mark the
boundaries of the 68\% containment range for shallow (a=0.5) and Bohmian (a=1)
energy scaling of the mean free path, respectively, both for a low GRB rate of
$0.5\ {\rm Myr}^{-1}$. For comparison, the black line is for a=0.5, but high GRB
rate $5\ {\rm Myr}^{-1}$.
\label{f-aniso2dir}}
\end{figure}

Figure \ref{f-aniso2} shows the anisotropy signal for the parameters of case B
(see Equation \ref{eq-mfp}). For Bohm-like scaling of the mean free path (case A),
the anisotropy is the same at $10^{17}$~eV but has a steeper energy dependence.
The current experimental limits on $\delta_{\rm tot}$ set by the Auger
Observatory are $\delta_{\rm tot}\le$0.01 in the [0.4,4] EeV range and
$\delta_{\rm tot}\le 0.05$ in the [4,40] EeV range
\citep{2008ICRC....4..175A,2009PrPNP..63..293B,bonino,2011APh....34..627P}.
To be noted from the figure is that
the anisotropy for protons above $10^{17}$~eV is high compared with the Auger
limits, in particular for the more realistic disk geometry. Note that the average
anisotropy (dashed and solid lines) is independent of the GRB rate.
A substantial contribution of heavy nuclei, a small mean free path,
or extragalactic UHECR would be needed to satisfy the anisotropy
limit. Figure \ref{f-aniso2dir} displays the variation in the anisotropy
direction, suggesting
that one may underestimate the anisotropy if one only considers
its projection on the anticenter direction,
fairly independent of the choice of parameters.

\begin{figure}
\includegraphics[width=0.7\columnwidth, keepaspectratio]{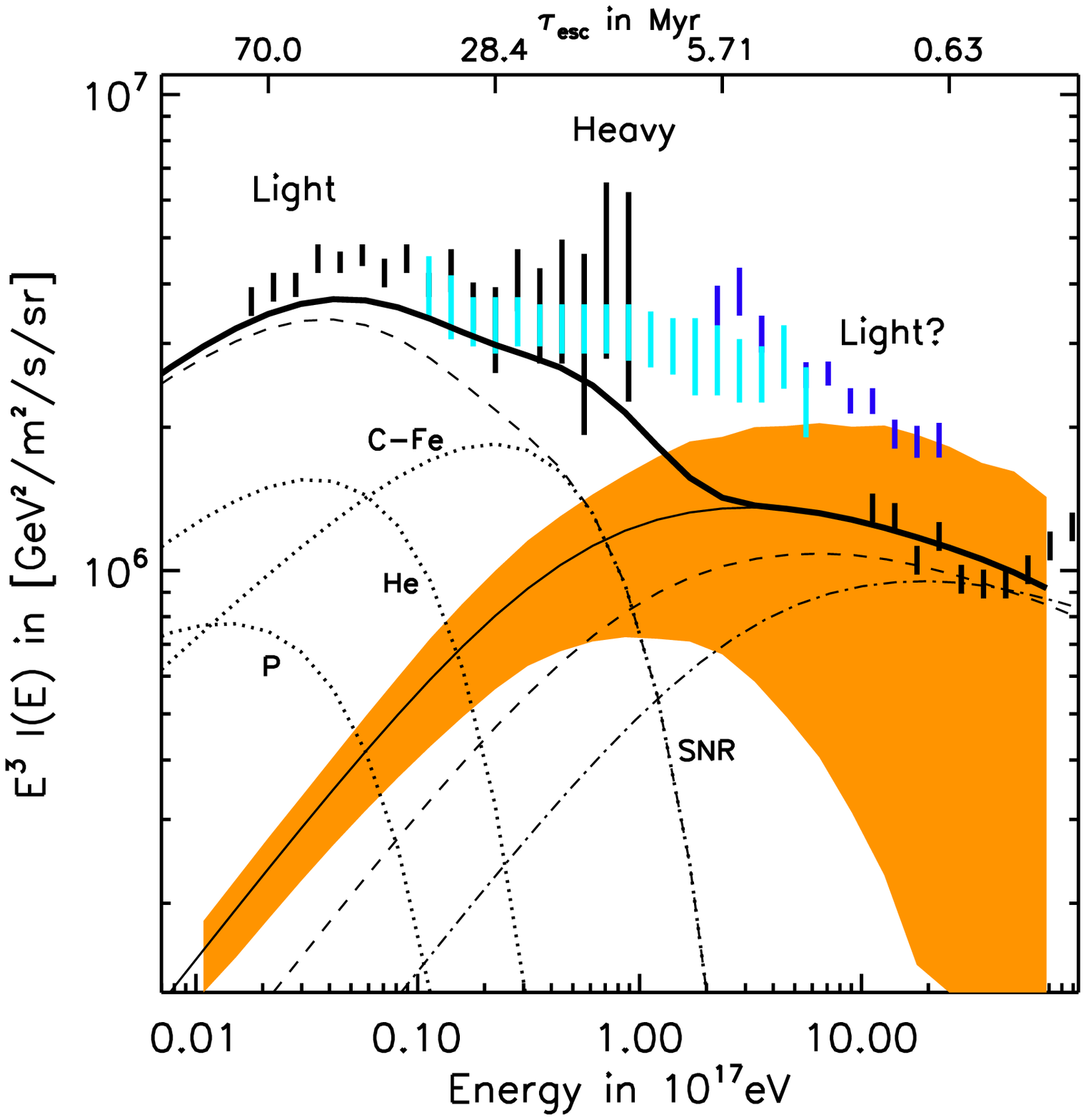}
\caption{Example of model spectra for cosmic-ray protons, helium,
and carbon nuclei, including the 68\% variation range in the case of protons.
The solid line
denotes the time-averaged spectrum of protons, the dashed line displays the same
for helium, and the dash-dotted line is for carbon.
The mean free path follows equation \ref{mfp}, and the corresponding escape timescales are indicated at the top axis. Also shown are the spectra measured with KASCADE-Grande below 0.5 EeV,
Auger above 1 EeV, and HiRes in between, together with labels indicating the composition. The offset
between spectra from different experiments is likely due
to errors in the absolute energy scale. For comparison, we also display correspondingly labelled
simple mock spectra of 
protons, helium, and C-Fe-group particles possibly produced in SNR. The thick triple-dot-dashed
line indicates the total of protons from GRB and all particles from SNR. 
\label{f-model}}
\end{figure}
\begin{figure}
\includegraphics[width=0.7\columnwidth, keepaspectratio]{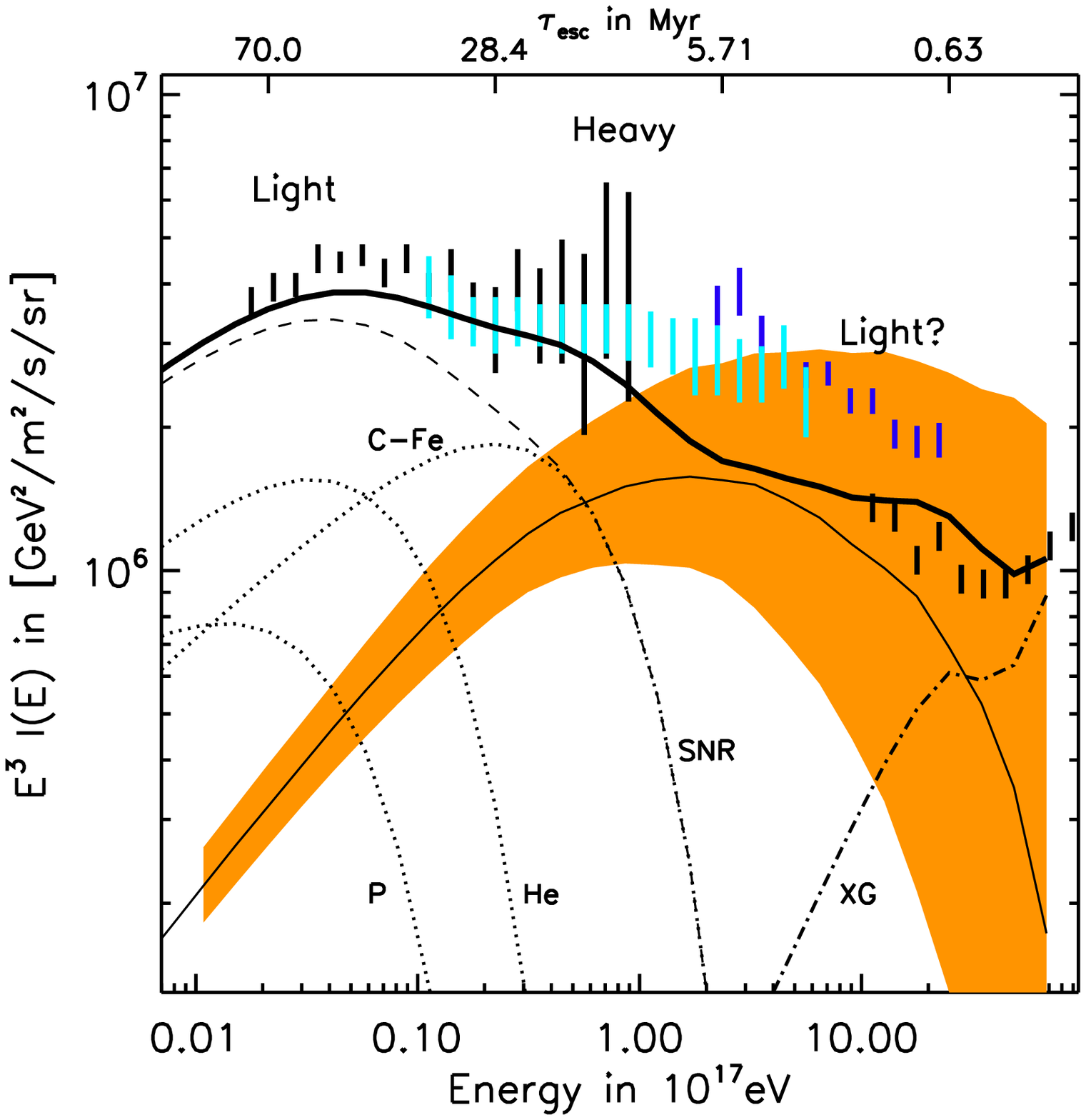}
\caption{Example of model spectra for cosmic-ray protons in the presence of an extragalactic
proton component. The solid line
denotes the median spectrum of protons, and the dash-dotted line displays the extragalactic proton
spectrum. Otherwise, everything is as in Figure \ref{f-model}.
\label{f-model+xg}}
\end{figure}

\section{Combined analysis}\label{sec3}
We now try to construct a model that reproduces the spectrum of
cosmic rays between $4\cdot 10^{16}$~eV and $4\cdot 10^{18}$~eV together with the
anisotropy limits and the composition. We use data of the Kascade-Grande
collaboration \citep{2009APh....31...86A,2010arXiv1009.4716A}, ranging from 2 PeV to 0.5 EeV,
HiRes monocular data between 0.2 EeV and 3 EeV
\citep{2005PhLB..619..271A},
and the Auger collaboration \citep{2010PhLB..685..239P} above 1 EeV.

\begin{figure}
\includegraphics[width=0.7\columnwidth, keepaspectratio]{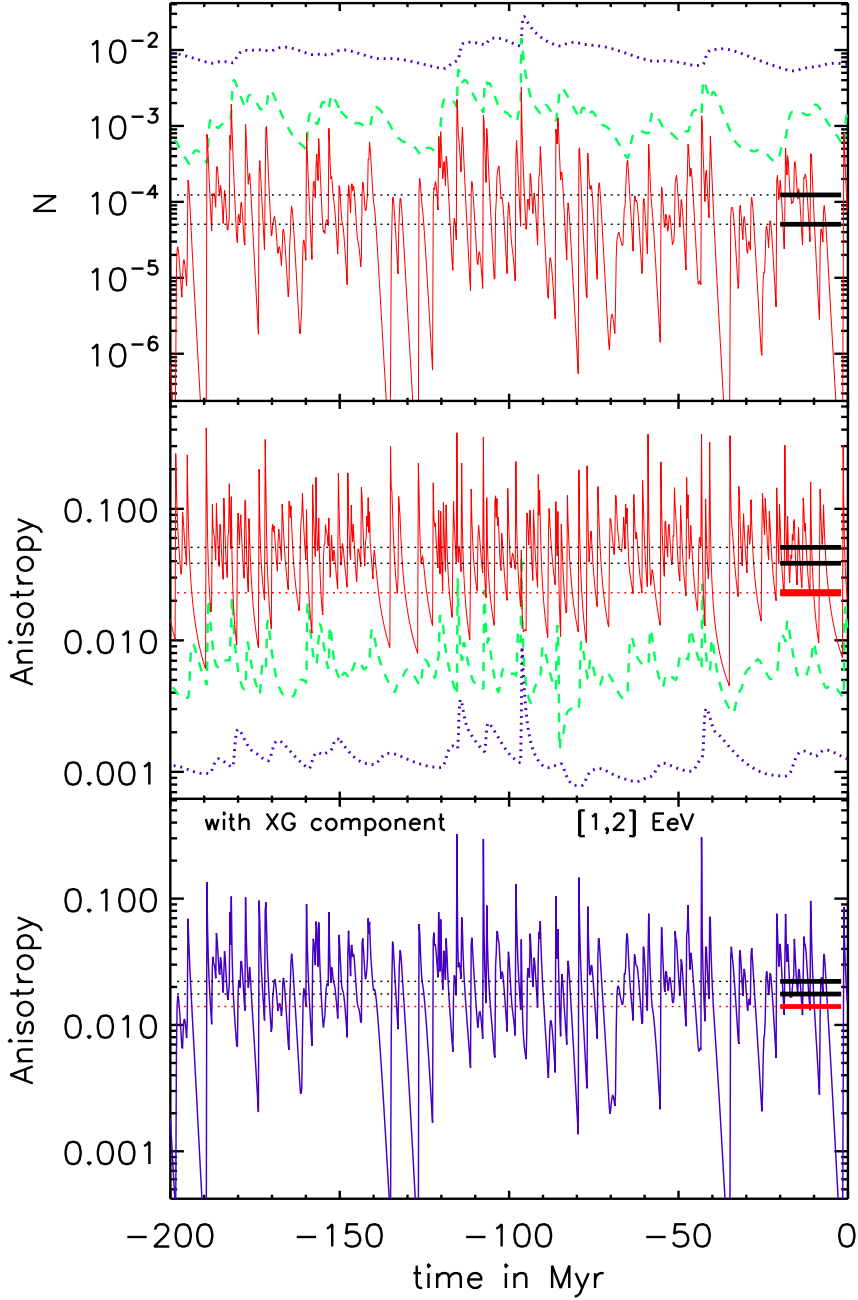}
\caption{{\it Top panel:} 200-Myr lightcurve of the local proton flux
(cf. Figure \ref{f-model}) in arbitrary units at 3
different escape times relative to the inverse source rate of 1 Myr: The dotted (blue) line is for 
$\tau_{\rm esc}= 20$ Myr , the dashed (cyan) line for $\tau_{\rm esc}= 4.5 $ Myr, and the solid
(red) line for $\tau_{\rm esc}= 0.5 $ Myr.
The two black bars on the right of the panels, extended as thin dashed lines,
indicate the median and the average of the light curves at 
$\tau_{\rm esc}= 5 \cdot 10^5$ yr, where the lower bar always indicates the median.
{\it Middle panel:}
Temporal variation of the anisotropy at the same escape times. For the diffusion parameters
chosen in the text, where the mean free path is chosen conservatively
as the gyroradius for the two highest energies, the dotted (blue) line is for
$0.024$~EeV, the dashed (cyan) line for $0.24$~EeV, and the solid(red) line for $2.4$~EeV.
The red bar on the right indicates the 99\% upper limit set by Auger.  
{\it Bottom panel:}
Temporal variation of the anisotropy for the model including an extragalactic proton
component (cf. Figure \ref{f-model+xg}), here only between $1$~EeV and $2$~EeV.
\label{lc-compd}}
\end{figure}

At $10^{18}$~eV the anisotropy is low, $\delta \lesssim 0.01$ (the 99\% upper limit is
0.014) \citep{2011APh....34..627P}, and the
composition is light, though not necessarily dominated by protons \citep{2010PhRvL.104i1101A}.
Figure \ref{f-aniso2} suggests that the mean free path must be
relatively small to accommodate the anisotropy limits for light nuclei
such as helium. Cosmic-ray propagation models can reproduce the
composition and particle spectra in the GeV--TeV band for a shallow
energy dependence of the diffusion coefficient ($D\propto E^{0.33}$),
if stochastic reacceleration is included \citep{sm98,mau01,jon01}.
If we were to extrapolate their
results, the proton mean free path at $10^{17}$~eV would be
$\sim 500$~pc, i.e. larger than what we have assumed above. Obviously,
we cannot permit the extrapolation to hold, if Galactic GRB are to
account for the observed UHECRs in the $10^{18}$~eV energy band. The
mean free path must be smaller than the extrapolation, but how much
smaller, in particular how small compared with the particle Larmor radius?

Figure \ref{f-model} shows spectra for a possible model configuration,
where for simplicity we display only spectra for protons, helium, and carbon as proxies
for light and heavy nuclei, respectively. The GRB rate is set to
$P(t)=1\ {\rm Myr}^{-1}$ and the source spectral index is $s=2.1$. The mean
free path transitions from a shallow energy dependence to Bohmian
scaling ($\propto E$) as the particle energy increases,
\begin{equation}
\lambda_{\rm mfp}=\lambda_0\,E^{0.3}\,\left[1+\frac{E}{60\,\rm PeV}\right]^{0.7}
\label{mfp}
\end{equation}
where $\lambda_0$ is chosen so a proton has a mean free path
that is a certain multiple (unity in Figure \ref{f-model}) of $11$~pc at
$10^{17}$~eV, its Larmor radius in a 10-$\mu$G magnetic field. The top
axis of Figure \ref{f-model} indicates the escape time at a few energies. The
spectrum below $10^{17}$~eV is far below observed values to accommodate
other Galactic sources of cosmic rays, such as SNR or PWN, for which we plot simple power
laws with exponential cut-off at $Z\cdot (3\ {\rm PeV})$ to guide the eye, based on
\citet{2010ApJ...714L..89A}. For that we assume the averages $<A>=25$ and $<Z>=12.5$ for the C-Fe
group and the following power-law indices: $s=2.66$ for protons, $s=2.58$ for helium, and 
$s=2.56$ for C-Fe. It is apparent that the energy scales of HiRes and Auger are discrepant. To fully fit the data we need to increase the normalization of the GRB-particle spectrum and either disregard
the Auger data or, alternatively,
distrust HiRes and require a fluctuation in the local spectrum that makes the GRB-produced protons 
somewhat softer than the average spectrum plotted in Figure \ref{f-model}. Such a deviation would be
common, because the fluctuation
amplitude at escape times below 1 Myr is large for the GRB rate used here,
one per Galaxy per Myr.

The required spectral shape above $1$~EeV depends on the properties of the extragalactic component that dominates above the ankle. \citet{2006PhRvD..74d3005B} have demonstrated that the effect of 
the pair-production threshold on an extragalactic component composed solely of protons can match the ankle fairly well. In Figure \ref{f-model+xg} we add such a component using a simple approximation of the pair-production loss rate ($\dot E\propto E\,\Theta(E-E_{\rm thr.})$, assuming no evolution,
i.e. $m=0$ in \citep{2006PhRvD..74d3005B}, and integrating out to a redshift $z_{\rm max}=5$). The evolution and redshift cut-off can in principle be constrained using the neutrino and gamma-ray backgrounds \citep[e.g.][]{wang2011}, but such an endeavor is beyond the scope of this paper.
In calculating the anisotropy
(cf. Figure \ref{lc-compd}), we assume the extragalactic component to be perfectly isotropic. The motion of the Sun and the Galaxy relative to the CMB frame suggests the presence of a
Compton-Getting anisotropy of amplitude $\sim 0.5\%$ \citep{2006PhLB..640..225K}, but it is unclear 
to what degree scattering in the Galaxy can reduce the observed anisotropy. Assuming no anisotropy 
in the extragalactic component therefore constitutes a best-case scenario (in the sense of diluting the anisotropy) that is likely too optimistic.
Note that we slightly rescaled the component from galactic GRB and plot the median instead of the time-average.

The average spectrum mainly depends on the absolute value of the mean free
path at high energies. We can calculate the cosmic-ray source power required
to sustain the observed flux of UHECRs at $10^{18}$~eV as function of
the mean free path. For the parameters used here, and an injection spectrum $\propto E^{-2.1}$ extending from
the GeV band to the highest energies, fitting the observed flux of UHECRs at
$10^{18}$~eV requires the source power
\begin{equation}
P_{\rm CR}=P(t)\,\int_{1\ \rm GeV} dE\ E\,Q_0\,E^{-s}
\simeq \frac{\lambda_{\rm mfp}}{r_L}\ (10^{37}\ {\rm erg/s})
\end{equation}
The source power in the energy interval $[10^{17},10^{18}]$~eV alone is
\begin{equation}
P_{\rm EeV}
\simeq \frac{\lambda_{\rm mfp}}{r_L}\ (3.5\cdot 10^{35}\ {\rm erg/s}) ,
\end{equation}
which is about $10^{-5}$ of the total CR luminosity of the Galaxy.
By way of comparison, the source intensity of alternative sources to SNR,
if they are to supply all Galactic CR at energies above the Lagage-Cesarsky limit
of $E_{\rm LC} = 10^{14.5}$~eV for SNR, must be the Galactic luminosity of CR above
these energies, i.e.
\begin{equation}
P_{\rm SNR}\approx 10^{(-0.7)(5.5)}\,L_{\rm GeV}\,
\frac{\tau_{\rm esc}({\rm GeV})}{\tau_{\rm esc}\,(10^{5.5}\,\rm GeV)}
\simeq (10^{36.5}\ {\rm erg/s})\,\frac{\tau_{\rm esc}({\rm GeV})}{\tau_{\rm esc}\,(10^{5.5}\,\rm GeV)} .
\end{equation}
Assuming that
$\tau_{\rm esc}({\rm GeV})/\tau_{\rm esc}(10^{5.5} {\rm GeV})\ge 30$, this is
about $10^{38}$\, erg/s
above $10^{5.5}$~GeV. On the other hand, supernovae may supply the observed CR all the way to $10^{16}$ eV if those at the highest energies are heavy nuclei.

\begin{figure}
\includegraphics[width=0.7\columnwidth, keepaspectratio]{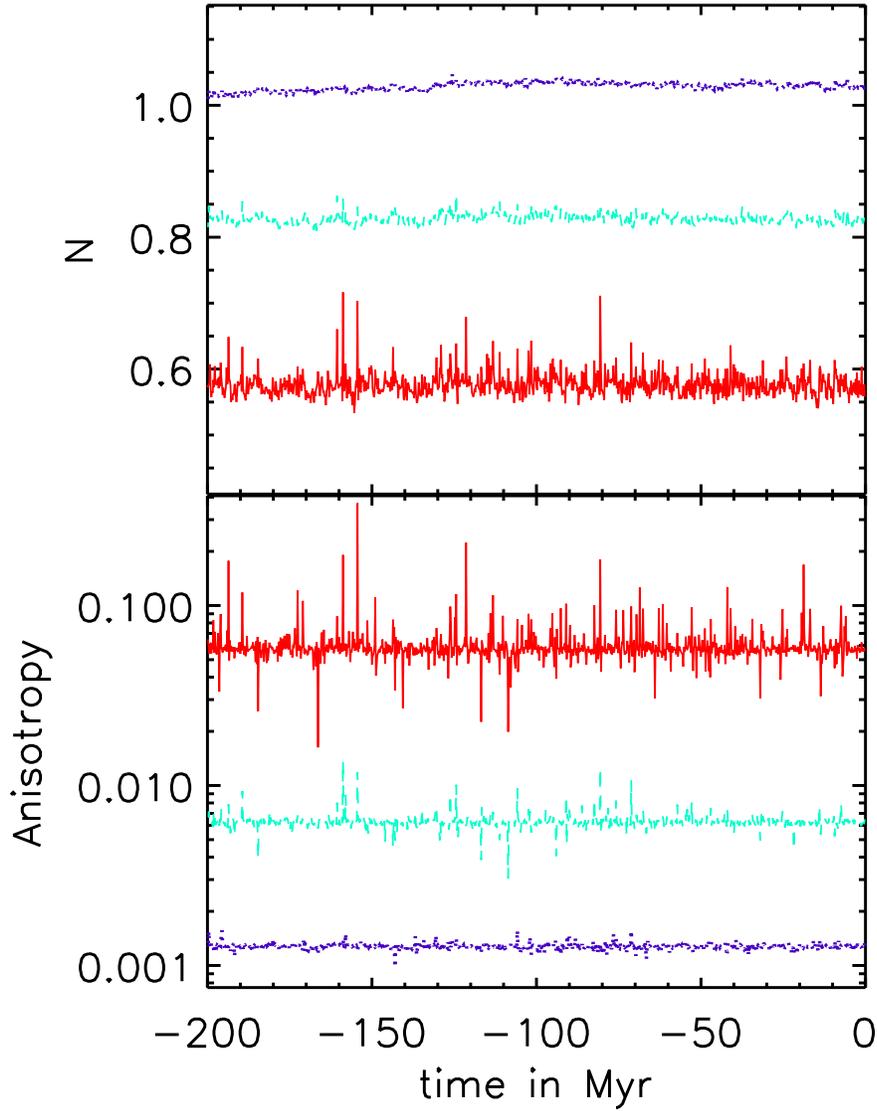}
\caption{{\it Top panel:} 200-Myr lightcurve of the local proton flux at the 3
different escape times denoted in Figure \ref{lc-compd}, assuming the protons originate
in SNR with the parameters described in the text.  {\it Bottom panel:}
Temporal variation of the anisotropy at the same energies.
\label{lc-sn}}
\end{figure}
What Figures \ref{f-model} and \ref{f-model+xg} serve to demonstrate is that a GRB produced
cosmic-ray component can fill the gap between the contributions of ordinary galactic cosmic-ray sources and the super-ankle region likely supplied by extragalactic sources. The discreteness
fluctuations are typically larger than experimental uncertainties. But what of the anisotropy
and its variation?

The lightcurve shown in the top panel of Figure \ref{lc-compd} shows that
besides the increase in fluctuation amplitude, the fluctuation timescale
decreases with particle energy (or more precisely escape probability). For the parameters used here, the escape time at
$2.4$~EeV is about $5\cdot 10^5$~yr. The flux variations can be compared with the
temporal behaviour of the anisotropy, shown in the middle panel of Figure
\ref{lc-compd}. Already at $0.24$~EeV (the dashed, cyan line) the anisotropy
hovers around 1\%. A clear correlation between flux and anisotropy is not
discernible at higher energies, although at $0.024$~EeV some structures in flux and
anisotropy coincide. However, other spikes in flux or anisotropy observed at $0.024$~EeV
have no counterpart at higher energies, and vice versa. To be noted from
the figures is that significant excursions to low anisotropy are rare. The minimum anisotropy is
$0.0045$, whereas the minimum flux is $8\cdot 10^{-9}$. 

Figure \ref{lc-compd} also indicates the median and the time average of the flux and anisotropy
at $2.4$~EeV. The average is what we would expect to observe if the number of UHECR sources
were much larger. In contrast, the median is a typical result in the case of intermittent sources.
To be noted from the figure is that both the average and the median at $2.4$~EeV are considerably in excess of the Auger limits. Nevertheless, 
the anisotropy falls below the 99\% upper limit of 2.3\% \citep[For the energy band between 
$2$ and $4$~EeV][]{2011APh....34..627P} during about 25\% of the time.
The bottom panel of Figure \ref{lc-compd} shows the anisotropy variation in the energy band
$\left[1,2\right]$~EeV for the case
including an extragalactic proton component as in Figure \ref{f-model+xg},
which we assume to be perfectly isotropic. The average anisotropy is slightly reduced, but we note that the anisotropy becomes very small during lulls, essentially because during these times the extragalactic component dominates the spectrum. Still, the anisotropy falls below the 
99\% upper limit of 1.4\% only during 40\% of the time.

During the long lulls the intensity
dips as low as {\it three or four} orders of magnitude below the time average! This possibility
would suggests an anthropic scenario whereby we live in highly unusual times, and that the
present cosmic-ray intensity is, perhaps, far below usual conditions, which for some (possibly obscure) reason would be hostile
to intelligent life and/or advanced civilization. Note that the extreme versions of this scenario, in which the present UHECR luminosity of the Galaxy is $\sim 10^{-4}$ below the average (and possibly an even larger factor below the most recent maximum), would require an average (or peak) luminosity above 100 GeV or so that compares with (or exceeds) the present day value, and hence an extremely flat Galactic source spectrum. This could possibly be tested or constrained by  abundance measurements of terrestrial cosmogenic nuclei (TCN).
It should be remembered that the hypothesis of sharp time dependence applies only to the CR that escape rapidly enough, and the anisotropy bounds motivate this only at the highest energies. Thus, the TCN production rate as a whole is not necessarily strongly affected by the hypothesis, so TCN  anomalies are not firmly predicted by it. On the other hand, the production of ionizing particles at sea level per unit primary energy is an increasing function of the latter (assuming shower maximum occurs above sea level).  Thus, for a perfectly flat spectrum (equal energy flux per decade of primary energy), most of the ionizing secondary flux at sea level, which might possibly be of astrobiological relevance,  would come from CR at ultrahigh energies.
Further discussion of an anthropic scenario is beyond the scope of this paper, but will be the subject of future work. It is mentioned here to justify consideration of a scenario that would otherwise seem  {\it a priori} improbable, but accounts for the remarkably low limits on anisotropy at ultrahigh energy. 

\begin{figure}
\includegraphics[width=0.7\columnwidth, keepaspectratio]{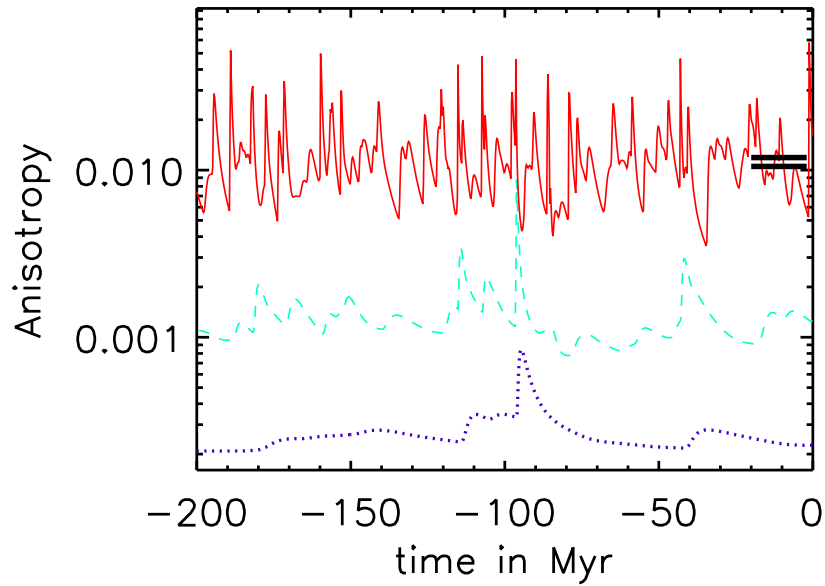}
\caption{Temporal variation of the anisotropy as in the bottom panel of Figure
\ref{lc-compd}, but for a smaller mean free path a
factor 5 smaller than assumed there, i.e.
$\lambda_{\rm mfp} =0.2\,r_L$ at high energies. Again the average and the median at
$2.4$~EeV are indicated which are both close to $1\%$, approximately at the level of the Auger upper limits.
\label{lc-anisod.2}}
\end{figure}

Also to be noted is that the high anisotropy is unavoidable in scenarios involving UHECR acceleration
in SNRs, because intermittency is not an issue and there is little uncertainty concerning the
spatial distribution of SNRs in the Galaxy. Lightcurves for an SNR origin are shown in Figure
\ref{lc-sn}, for which we assumed a supernova rate of one per 50 years. The rare spikes in
intensity are caused by very nearby supernovae, for which their spatial and temporal evolution
becomes an issue. Since we maintain the duration of the particle injection per source,
$10^{-2}$~Myr, the spikes are likely clipped by temporal smoothing.

The anisotropy can in principle be reduced by choosing a smaller mean free path.
Figure \ref{lc-anisod.2} shows the anisotropy for protons and a mean free path
following equation \ref{mfp}, but with $\lambda_0$ chosen so that
at high energy $\lambda_{\rm mfp} =0.2\,r_L$. The Auger limits, $\delta \lesssim
0.01$ at $1$~EeV, can be satisfied with protons. The margin is not large, though,
and a few more years of Auger operation should permit detecting an anisotropy
signal, even for an unusually small mean free path.

\section{Summary and discussion}
We have calculated the time-dependent transport of UHECR in the Galaxy, assuming
it can be described as isotropic diffusion. We have in particular investigated the
question, whether or not Galactic long GRB can provide the observed particles up to the ankle,
which is a natural transition point to extragalactic particles. In an earlier paper of the
same series (Paper II) it was demonstrated that if extragalactic GRB are to produce
UHECR above the ankle, Galactic GR should contribute a large flux of sub-ankle particles.
Our present findings can be summarized as follows:

\begin{itemize}
\item With the assumption of spherical symmetry, one overestimates
the UHECR flux by a factor of a few, and
likewise underestimates the anisotropy.
\item Intermittency becomes serious if the mean free path for scattering exceeds 100~pc, unless
the source rate is much higher than 1 per Myr. On average, Galactic long GRB  need to contribute only 
about $10^{37}$~erg/s in accelerated particles to fully account for the observed particle
flux at $10^{18}$~eV, assuming a Bohmian mean free path at this energy.
\item UHECR from Galactic long GRB can meet the observational limits on anisotropy only if the
mean free path for scattering is sufficiently small. Contributing the observed sub-ankle particles
(at $10^{18}$~eV) requires Bohmian diffusion if the UHECR are as heavy as carbon. A light composition
such as protons or helium requires sub-Bohmian diffusion, which is a highly unlikely situation for
isotropic diffusion. We have not investigated the effects of a Galactic guiding field that may modify
the probability of escape from the Galactic disk.
\item Auger data suggest that at $10^{18}$~eV the composition is indeed light, thus posing a problem
for the notion that Galactic GRB (or any other source class with similar population statistic) produce
the observed UHECR up to the ankle. (This measurement is not undisputed, though, for the KASCADE-Grande
collaboration has just published their analysis results which seem to favor a relatively heavy
composition up to nearly $10^{18}$~eV \citep{2010arXiv1009.4716A}.)
\item The UHECR composition is a very critical constraint, but its measurement
is subject to considerable systematic uncertainties arising from its dependence on models for the
development of air showers. It is imperative that measures be taken to better understand
the air-shower physics near $10^{18}$~eV.
\item Much of the UHECR anisotropy arises from the expected location of long GRB in the inner Galaxy.
Observations of GRB host galaxies suggest that regions of low metallicity and high star formation
may be the preferred sites of long GRB \citep{2010AJ....140.1557L,2011arXiv1101.4418L}, which
may skew the galactocentric distribution of long GRB toward the outer Galaxy.
As there is no power problem with Galactic GRB, it may be worthwhile to also consider short GRB.
They provide supposedly less power as a population, but they may have a very extended
spatial distribution in the Galaxy, thus reducing the anisotropy \citep{2010ApJ...722.1946B}.
The anisotropy arising from intermittency would remain in both cases and is the subject
of a forthcoming publication.
\item Our study can be applied with little change to the case of an UHECR origin
in supernova remnants, assuming very efficient magnetic-field amplification can increase their ability to accelerate particles to energies significantly higher than $1$~PeV \citep[e.g.][]{2001MNRAS.321..433B}. The spatial distribution in the Galaxy of long GRB and SNR can be expected to be similar, and therefore the average anisotropy is the same for both long GRB and SNR. We have verified that intermittency would be negligible in case of a SNR origin (unless of course the UHECR come only from a rare subclass of supernovae) and therefore problems in simultaneously matching the composition and the anisotropy limits of UHECR are inescapable for SNR.
\end{itemize}

\acknowledgements
Support from the Israel Science Foundation,  the Israel-U.S. Binational Science Foundation, and the
Joan and Robert Arnow Chair of Theoretical Astrophysics  is
gratefully acknowledged.

\end{document}